\documentclass[twocolumn,preprintnumbers,amsmath,amssymb,aps,prb,longbibliography]{revtex4-1}
\usepackage{graphicx}
\usepackage{color}
\usepackage{dcolumn}
\usepackage{bm}
\usepackage{hyperref}
\usepackage{mathrsfs}
\begin{document}

\title{Controlled Skyrmion Ratchet in Linear Protrusion Defects}
\author{J. C. Bellizotti Souza$^{1}$,
            N. P. Vizarim$^{2}$, 
            C. J. O. Reichhardt$^3$, 
            C. Reichhardt$^3$
            and P. A. Venegas$^2$}
            
\affiliation{$^1$ POSMAT - Programa de P\'os-Gradua\c{c}\~ao em Ci\^encia e Tecnologia de Materiais, Faculdade de Ci\^encias, Universidade Estadual Paulista - UNESP, Bauru, SP, CP 473, 17033-360, Brazil\\
$^2$ Departamento de F\'isica, Faculdade de Ci\^encias, Unesp-Universidade Estadual Paulista, CP 473, 17033-360 Bauru, SP, Brazil\\
  $^3$ Theoretical Division and Center for Nonlinear Studies, Los Alamos National Laboratory, Los Alamos, New Mexico 87545, USA}

\date{\today}

\begin{abstract}
Using atomistic simulations, we investigate the dynamical behavior
of a single skyrmion interacting with an asymmetric linear protrusion
array under external ac driving.
When the ac drive is applied along the $x$ direction, the skyrmion moves along
the hard direction of the substrate asymmetry in three
phases: a pinned phase with localized skyrmion orbits,
a constant velocity phase where the orbits become delocalized,
and a reentrant pinned phase with larger
localized orbits. We measure the dependence of the skyrmion
velocity on the frequency and amplitude of the ac drive. All three
phases appear for all frequency values, and in the constant velocity
phase the skyrmion velocity depends only on the frequency and not
on the amplitude of the ac drive.
When ac driving is applied in the $y$ direction, 
the skyrmion moves along the easy direction of the substrate asymmetry
and exhibits the same three phases
as for $x$ direction driving along with a fourth phase which,
at high driving frequencies,
consists of a series of constant velocity phases,
each with different average skyrmion velocities.
For low frequencies, the
constant velocity phase is lost
and the skyrmion speed increases linearly with increasing
ac drive amplitude due to a Magnus boost effect.
Our findings suggest new ways
to create reliable data transport for spintronic
devices using skyrmions as information carriers,
where the skyrmion direction and speed can be controlled
by varying only the ac drive amplitude and frequency.
\end{abstract}

\maketitle

\section{Introduction}
In the field of data storage,
there is great interest in
discovering novel approaches
that demand lower energy consumption
and offer increased data density.
The rapid increase in cloud-based storage has resulted in substantial energy consumption \cite{negru_energy_2013},
a concern that can be mitigated through new data storage techniques capable of 
enhancing data density and data transfer efficiency.
One promising direction is to employ stable topological stable
objects with reduced dimensions whose transport can be
controlled precisely.
Recently magnetic skyrmions, which are topologically stable spin textures
\cite{nagaosa_topological_2013,je_direct_2020} of size
ranging from a few nanometers up to a micrometer \cite{wang_theory_2018},
have been observed experimentally
in chiral ferromagnetic thin films and bulk crystals
\cite{muhlbauer_skyrmion_2009,yu_real-space_2010,pfleiderer_skyrmion_2010,munzer_skyrmion_2010}.
Skyrmions are considered one of the most promising candidates for future spintronic devices
\cite{nagaosa_topological_2013,everschor-sitte_perspective_2018,fert_skyrmions_2013}
due to their reduced size and stability even at room temperatures \cite{fert_skyrmions_2013,fert_magnetic_2017}, as well as their ability to
be transported via the application
of a spin-polarized current. The
current density necessary to drive a skyrmion
is over five orders of magnitude smaller than that required to
move magnetic domain walls \cite{schulz_emergent_2012,jonietz_spin_2010}, and
as a result,
skyrmion-based devices
could provide more reliable and energy-efficient
high density data storage.
Some of the proposed skyrmion-based devices include
magnetic logic gates
\cite{luo_skyrmion_2021,shu_realization_2022,zhang_magnetic_2015-1},
diodes
\cite{shu_realization_2022,bellizotti_souza_magnus_2022,feng_skyrmion_2022,wang_magnetic_2020,song_skyrmion-based_2020,jung_magnetic_2021,zhao_ferromagnetic_2020}, and
transistors \cite{zhang_magnetic_2015}.
Magnetic skyrmions could also be used in
non-conventional computing, such as
audio classification artificial intelligence \cite{msiska_audio_2022}
and neuromorphic computing \cite{song_skyrmion-based_2020,li_magnetic_2017,li_magnetic_2021}.
In order to harness magnetic skyrmions effectively for spintronics devices
and unconventional computing, precise control over their motion is crucial,
and significant efforts have been devoted to achieve this goal
\cite{fert_skyrmions_2013,luo_reconfigurable_2018,zhang_laminar_2023,pfleiderer_surfaces_2011,wiesendanger_nanoscale_2016,kang_skyrmion-electronics_2016,zhang_skyrmion-electronics_2020}.

Skyrmions exhibit many similarities to other overdamped particles.
Both minimize their repulsive interactions by
forming a triangular lattice, can be set into
motion by external drives, and interact with
defects in the material \cite{olson_reichhardt_comparing_2014,reichhardt_depinning_2016}.
A distinguishing property of skyrmions is
the presence of a strong non-dissipative Magnus force,
which results in a very distinct dynamical behavior since
the Magnus force
creates a velocity component perpendicular
to the net force acting on the skyrmion.
The sign of this
perpendicular component depends on the skyrmion
winding number or topological charge
\cite{nagaosa_topological_2013,litzius_skyrmion_2017,iwasaki_universal_2013,jiang_direct_2017,lin_driven_2013,lin_particle_2013,zeissler_diameter-independent_2020}.
Due to the presence of the Magnus term, 
the skyrmion moves at what is known as the intrinsic skyrmion Hall angle
$\theta_\text{sk}^\text{int}$
\cite{nagaosa_topological_2013,litzius_skyrmion_2017,iwasaki_universal_2013,jiang_direct_2017,lin_driven_2013,lin_particle_2013}
with respect to an applied external drive.
The intrinsic skyrmion Hall angle is an important problem for technological applications,
since it can cause skyrmions to move toward the edge of the sample,
leading to annihilation events that limit how far the skyrmion can travel.
In order to prevent
this annihilation process, a precise
control of the skyrmion motion is required.
There is an increased effort to find new ways to control the skyrmion motion
and avoid the intrinsic skyrmion Hall angle problem. 
Some of the proposed methods include
the use of periodic pinning
\cite{reichhardt_quantized_2015,reichhardt_nonequilibrium_2018,feilhauer_controlled_2020,vizarim_directional_2021,vizarim_skyrmion_2020,reichhardt_commensuration_2022},
sample curvature \cite{carvalho-santos_skyrmion_2021,korniienko_effect_2020,yershov_curvature-induced_2022},
interface guided motion \cite{vizarim_guided_2021,zhang_edge-guided_2022},
ratchet effects \cite{reichhardt_magnus-induced_2015,souza_skyrmion_2021,chen_skyrmion_2019,gobel_skyrmion_2021},
temperature and magnetic gradients \cite{yanes_skyrmion_2019,zhang_manipulation_2018,everschor_rotating_2012,kong_dynamics_2013},
skyrmion-vortex coupling using heterostructures \cite{menezes_manipulation_2019,neto_mesoscale_2022},
skyrmion lattice compression \cite{zhang_structural_2022,bellizotti_souza_spontaneous_2023},
laminar flow of skyrmions \cite{zhang_laminar_2023},
soliton motion along skyrmion chains \cite{vizarim_soliton_2022,souza_soliton_2023}, and skyrmions interacting with chiral flowers \cite{zhang_chiral_2023}.

The interaction of overdamped particles
with asymmetric potentials under dc or ac driving
has been explored widely
\cite{wambaugh_superconducting_1999,vlasko-vlasov_jamming_2013,souza_clogging_2022,martinez_trapping_2020,reichhardt_magnus-induced_2015,gonzalez_transverse_2007,lu_reversible_2007,olson_reichhardt_vortex_2013,villegas_experimental_2005,yu_asymmetric_2007}.
For ac driving, the system may exhibit a ratchet effect in which
there is a net dc transport of the particle
that is generally along the easy flow direction of the potential.
One of the earliest skyrmion ratchet studies involved
a quasi-one-dimensional (1D) asymmetric substrate \cite{reichhardt_magnus-induced_2015}.
For ac drives applied parallel to the substrate asymmetry direction,
quantized net dc displacements of the skyrmion occurred,
while a Magnus induced transverse ratchet appeared
when the ac driving was perpendicular to the
substrate asymmetry direction.
In recent work, Souza \textit{et al.} \cite{souza_skyrmion_2021}
proposed a device where the motion of a single magnetic skyrmion
interacting with an asymmetric funnel array could be precisely
controlled, providing an information carrier.
With this geometry,
it was possible to produce ratcheting motion
of the skyrmion along both the easy and hard substrate asymmetry
directions;
however, the motion in the hard direction
had low efficiency.

\begin{figure}
\includegraphics[width=\columnwidth]{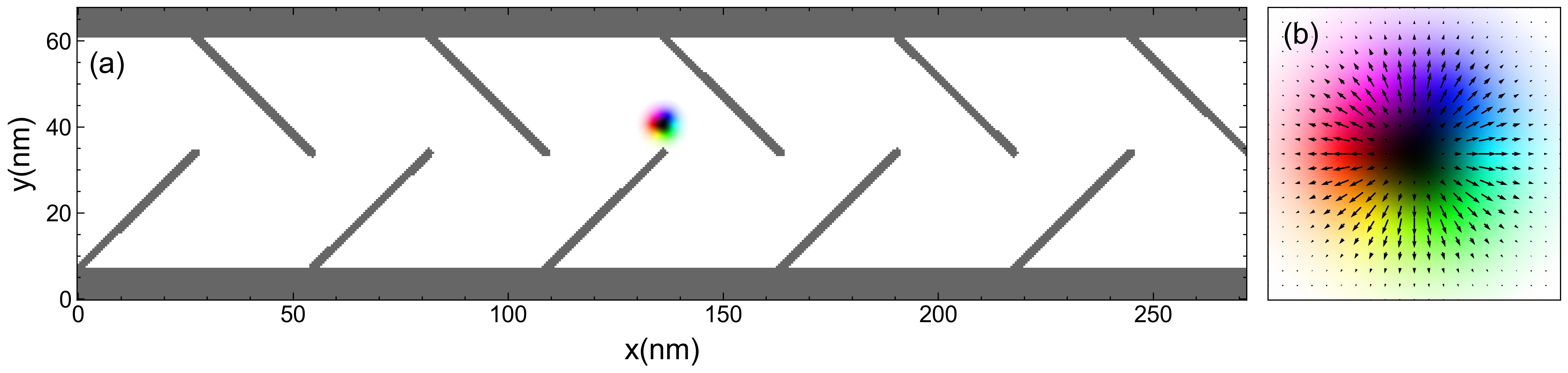}
\caption{(a) Illustration of the sample considered in this work with walls and 
  linear protrusions (gray) that are generated using very strong perpendicular
  magnetic anisotropy (PMA).
  The colored disk represents the skyrmion.
  (b) A blowup of the N{\' e}el skyrmion spin texture in
  the $x-y$ plane from panel (a),
  highlighting the magnetic moment components $m_x$ and $m_y$.}
\label{fig1}
\end{figure}

Another example of asymmetric arrays is the
linear protrusion array, shown in Fig. \ref{fig1},
that was previously used for type II superconducting
vortices by Wells \cite{wells_investigating_2017}.
In that work, an external dc driving force
was employed to induce a vortex diode effect.
The depinning threshold was low 
for driving currents applied along the easy flow direction and
high for driving along the hard flow direction.
Wells argued that ac driving
would induce a vortex ratchet effect.
Here, we perform a detailed analysis of the
behavior of magnetic skyrmions in a linear protrusion array,
focusing on the ratchet effects that arise under ac driving
with varied amplitude and frequency.

In this work we show that when the ac drive is applied along
the $x$ direction,
the skyrmion exhibits three dynamical phases:
(i) a pinned phase (PP), (ii) a constant velocity phase (CVP)
with motion along the hard substrate asymmetry direction, and
(iii) a reentrant pinned phase (RPP). 
Within the CVP, we find that the skyrmion velocity 
depends exclusively on the ac driving frequency.
In addition,
the interval of ac drive amplitudes for which the CVP
appears varies with the driving frequency.
When we apply the ac drive along the $y$ direction,
we uncover a more complex behavior featuring four phases: (i) PP,
(ii) CVP with motion along the easy substrate asymmetry direction,
(iii) RPP, and (iv) a mixture of CVP and a
linear increase of the skyrmion velocity with
increasing ac amplitude.
Each phase
appears within a well-defined range of ac amplitude and frequency.
The velocity characteristics in the CVP
state mirror those observed for ac driving along the $x$ direction, 
with the velocity being determined exclusively by the ac frequency. 
In the fourth phase, a succession of CVPs with distinct
average skyrmion velocities emerges at higher ac drive frequencies, while
for lower frequencies, the CVP dissipates
and the skyrmion velocity increases linearly with the ac drive amplitude.
A precise control of the skyrmion motion can be
achieved by tailoring the ranges of ac amplitude and frequency
in order to guide the skyrmion
in a desired fashion.

\section{Simulation}
The sample considered here is an ultrathin nanotrack of ferromagnetic material
that can host N{\' e}el skyrmions. The nanotrack has dimensions
of 272 nm $\times$ 64 nm and a magnetic field is applied
perpendicular to the film surface at $T=0$ K.
Periodic boundary conditions are applied
only along the $x$ direction. The sample contains
magnetic walls close to the sample edge and a linear protrusion array
of defects,
where the magnetic moments that compose the walls and the linear protrusion
array
have very strong perpendicular magnetic anisotropy (PMA).
We introduce the walls  in order to confine the skyrmions along the nanotrack.
Throughout this work we always consider the dynamics of a single skyrmion,
and the
initial skyrmion configuration is illustrated in Fig \ref{fig1}.

We use an atomistic model \cite{evans_atomistic_2018} for the simulations to investigate the skyrmion spin textures in detail.
The Hamiltonian governing the spin dynamics is given by
\cite{iwasaki_universal_2013,iwasaki_current-induced_2013,seki_skyrmions_2016}:

\begin{equation}\label{eq1}
  \begin{split}
    \mathscr{H}=-\sum_{i, j\in N}J_{i,j}\mathbf{m}_i\cdot\mathbf{m}_j
                -\sum_{i, j\in N}\mathbf{D}_{i,j}\cdot\left(\mathbf{m}_i\times\mathbf{m}_j\right)\\
                -\sum_i\mu\mathbf{H}\cdot\mathbf{m}_i
                -\sum_i K_1\left(\mathbf{m}_i\cdot\hat{\mathbf{z}}\right)^2
  \end{split}
\end{equation}

The first term on the right side is the exchange interaction between the nearest neighbors that compose the set $N$.
The underlying lattice is a square spin lattice with lattice constant $a$ and
exchange constant $J_{i,j}$ between spins $i$ and $j$.
The second term is the interfacial Dzyaloshinskii–Moriya
interaction, where $\mathbf{D}_{i,j}$ is the Dzyaloshinskii-Moriya vector between spins $i$ and $j$. 
The third term is the spin interaction
with an external applied magnetic field $\mathbf{H}$ known as the Zeeman interaction, where $\mu=\hbar\gamma$
is the magnetic moment magnitude and
$\gamma=1.76\times10^{11}$T$^{-1}$s$^{-1}$ is the electron
gyromagnetic ratio.
The last term is the sample easy-axis anisotropy, where $K_1$ is the anisotropy strength.
We model the set of spins $L$ that compose the linear protrusion array of
defects
as fixed magnetic
moments, $\mathbf{m}_{\in L}=-\hat{\mathbf{z}}$, and we note that
if we replace these fixed moments
by defects with very strong PMA values $K_L$, such as $K_{L}=5J$,
we obtain the same result that the motion of skyrmions through the
defects is prevented
\cite{zhang_particle-like_2022}.

The time evolution for the magnetic moments follows the LLG equation augmented
with the adiabatic spin transfer torque
\cite{seki_skyrmions_2016,slonczewski_dynamics_1972,gilbert_phenomenological_2004}:
\begin{equation}\label{eq2}
    \frac{\partial\mathbf{m}_i}{\partial t}=-\gamma\mathbf{m}_i\times\mathbf{H}^\text{eff}_i
                             +\alpha\mathbf{m}_i\times\frac{\partial\mathbf{m}_i}{\partial t}
                             +\frac{pa^3}{2e}\left(\mathbf{j}\cdot\nabla\right)\mathbf{m}_i \ .
\end{equation}
Here, $\gamma$ is the gyromagnetic ratio, 
$\mathbf{H}^\text{eff}_i=-\frac{1}{\hbar\gamma}\frac{\partial\mathscr{H}}{\partial \mathbf{m}_i}$ is the effective field
including all interactions in the Hamiltonian, $\alpha$ is
the phenomenological damping term introduced by Gilbert, and the last
term is the spin-transfer-torque (STT), where 
$p$ is the spin polarization,
$e$ the electron charge, and $\mathbf{j}$ the applied current density.
The STT current
includes the assumption that conduction electron spins are always parallel to
the magnetic moments $\mathbf{m}$ \cite{iwasaki_universal_2013,zang_dynamics_2011}.
We only consider the adiabatic contribution from the current interaction.
The non-adiabatic contribution can be neglected since it does not affect
the skyrmion dynamics significantly under small driving forces \cite{litzius_skyrmion_2017}.
We use an alternating STT current given by:

\begin{equation}\label{eq3}
    \mathbf{j}=j_x\cos\left(2\pi\omega t\right)\hat{\mathbf{x}}+ j_y\sin\left(2\pi\omega t\right)\hat{\mathbf{y}},
\end{equation}
where $\omega$ is the oscillating frequency, $t$ is the time, and $j_x$ and $j_y$ are the current amplitudes in the $x$ and $y$ directions, respectively.

The skyrmion velocity is calculated using the emergent electromagnetic fields \cite{schulz_emergent_2012,seki_skyrmions_2016}:
\begin{equation}\label{eq4}
  \begin{split}
    E^\text{em}_i=\frac{\hbar}{e}\mathbf{m}\cdot\left(\frac{\partial \mathbf{m}}{\partial i}\times\frac{\partial \mathbf{m}}{\partial t}\right)\;\;\;\;\\
    B^\text{em}_i=\frac{\hbar}{2e}\varepsilon_{ijk}\mathbf{m}\cdot\left(\frac{\partial \mathbf{m}}{\partial j}\times\frac{\partial \mathbf{m}}{\partial k}\right)
    \end{split}
\end{equation}
where $\varepsilon_{ijk}$ is the totally anti-symmetric tensor. The skyrmion
drift velocity is then computed using $\mathbf{E}^\text{em}=-\mathbf{v}_d\times\mathbf{B}^\text{em}$ \cite{schulz_emergent_2012,seki_skyrmions_2016}.

In our simulations we fix $\mu\mathbf{H}=0.5(D^2/J)(-\hat{\mathbf{z}})$,
$\alpha=0.3$, $p=-1.0$, and $a=0.5$nm, and we consider
$J=1$ meV, $D=0.18J$, and $K=0.02J$, which is similar to the magnetic parameters
found for MnSi \cite{iwasaki_universal_2013}. 
All our simulations start from the spin configuration shown in
Fig.~\ref{fig1}.
The numerical integration of Eq.~\ref{eq2}
is performed using a fourth order Runge-Kutta
method. For each value of the ac drive amplitude,
we calculate the time averaged skyrmion velocities
along the $x$ direction, $\left\langle v_x\right\rangle$,
over $3\times10^7$ timesteps to ensure a steady state.
We normalize the simulation time units to $t=(\hbar/J)\tau$ and the
current density units to $\mathbf{j}=(2eJ/\hbar a^2)\mathbf{j}'$,
where $\tau$ and $\mathbf{j}'$ are the normalized time and current, respectively.

\section{Ac drive along the $x$ direction}
We first consider ac driving applied
along the $x$ direction, so that $j_x \neq 0$ and $j_y=0$.
The ac drive frequency is $\omega=5.57\times10^7$ Hz and $j_x$
is in the range $0.19\times10^{10}\text{Am}^{-2}\leq j_x\leq 3.89\times10^{10}\text{Am}^{-2}$, which is
low enough that the skyrmion in the sample
does not annihilate.

\begin{figure}
\includegraphics[width=0.8\columnwidth]{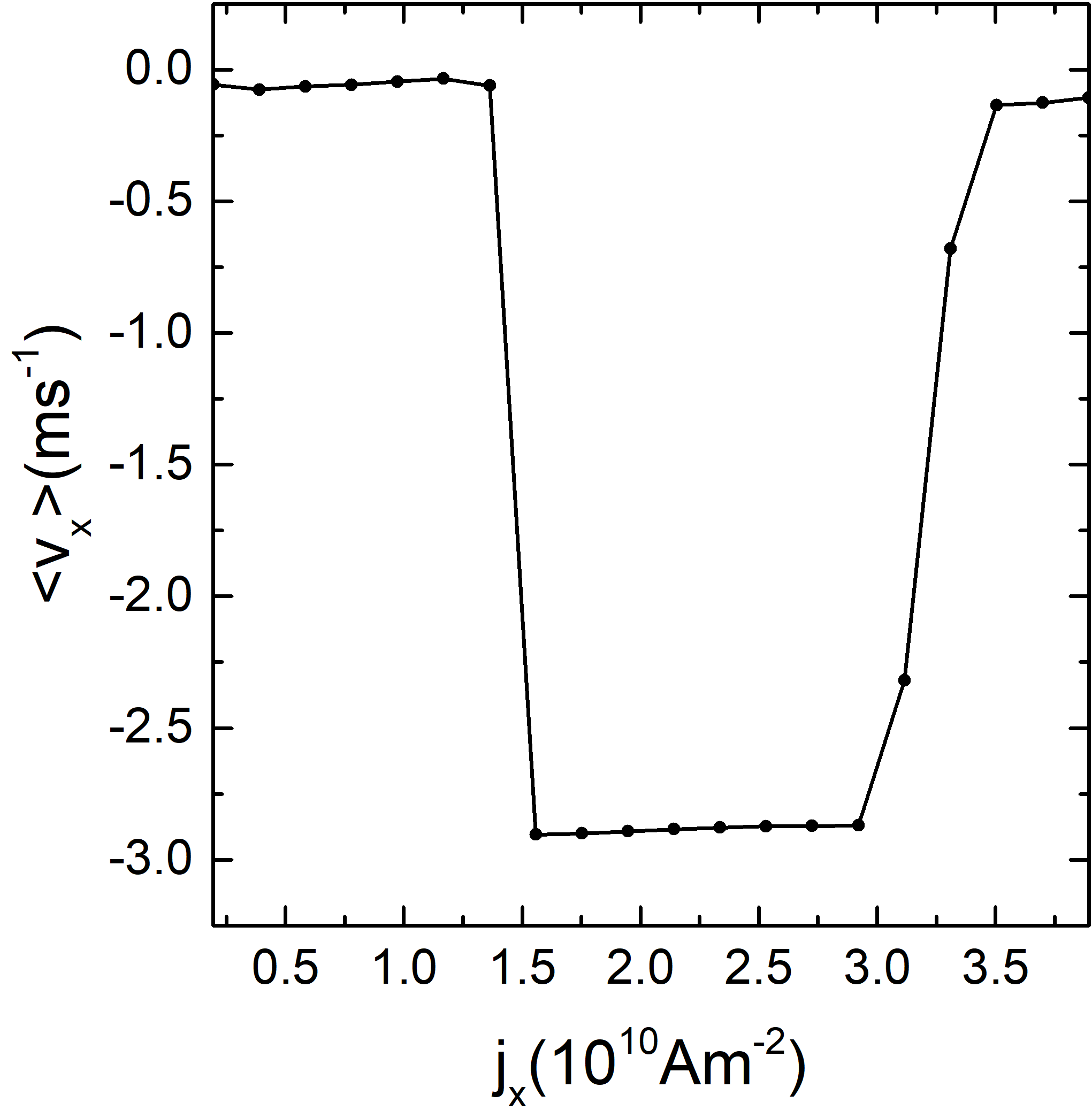}
\caption{$\left\langle v_x\right\rangle$ vs $j_x$ for the system from
  Fig.~\ref{fig1} with $x$ direction ac driving at
  $\omega=5.57\times10^7$ Hz.
  For $j_x \leq 1.36\times10^{10}\text{Am}^{-2}$,
  the skyrmion is in the pinned phase (PP), where the slight deviation of
  the velocity from zero is
  associated with transient motion that occurs before
  a stable localized orbit forms.
  The constant velocity phase (CVP) appears
 over the interval $1.5\times10^{10}\text{Am}^{-2}\leq j_x\leq 3.0\times10^{10}\text{Am}^{-2}$,
 where  $\left\langle v_x\right\rangle\approx -3 \text{ms}^{-1}$.
 Here the skyrmion moves in the negative $x$ direction.
    For larger values of $j_x$,
    the system enters a reentrant pinning phase (RPP),
    where the velocities are significantly reduced 
    but do not reach zero due to the transient motion.}
    \label{fig2}
\end{figure}

\begin{figure}
    \centering
    \includegraphics[width=\columnwidth]{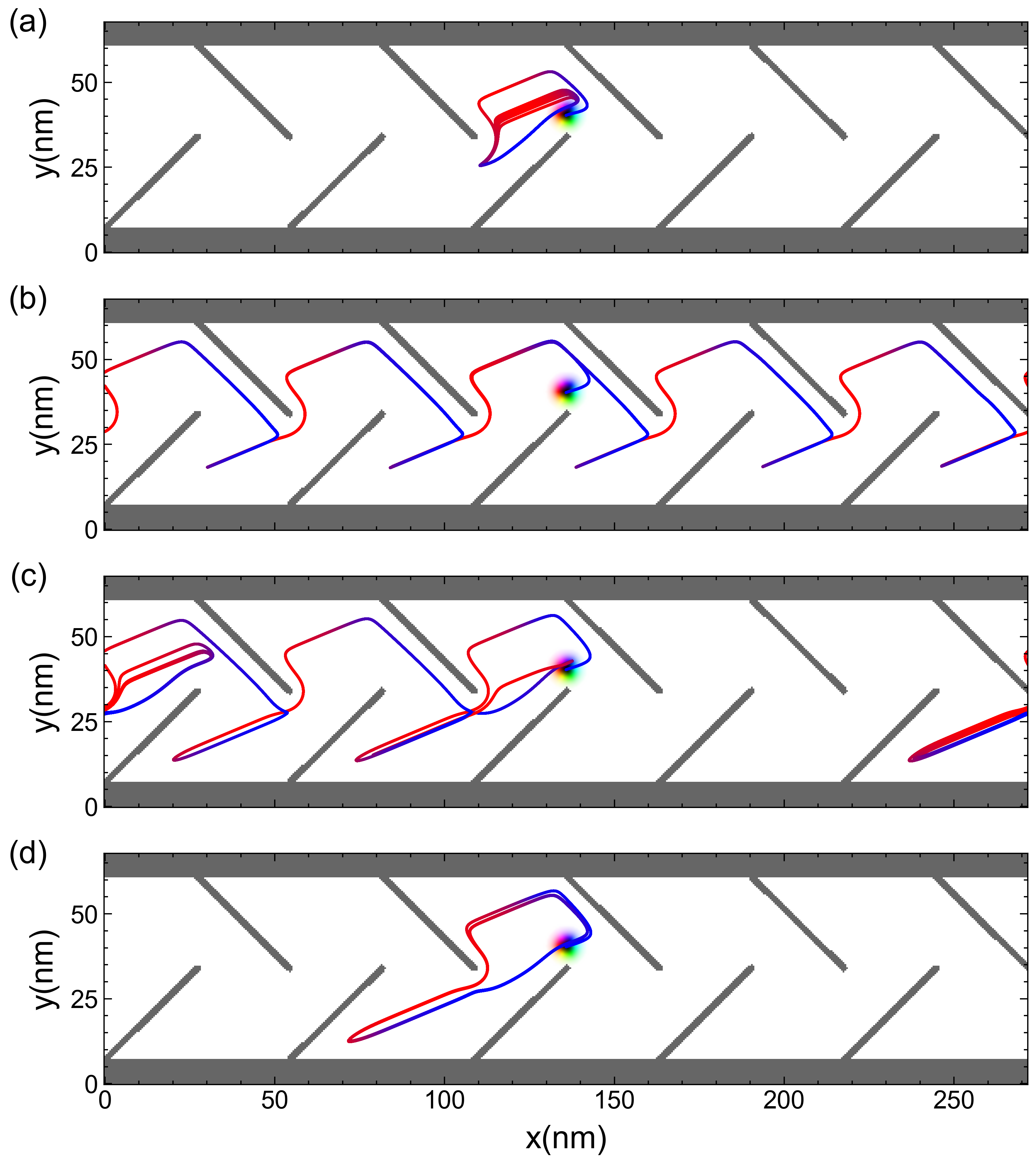}
    \caption{
      Illustration of skyrmion trajectories for the system in Fig.~\ref{fig2},
      with $x$ direction ac driving and $\omega=5.57\times10^7$ Hz.
      The skyrmion trajectory color indicates the phase of the ac drive
      cycle as a gradient from positive $x$ (blue) to negative $x$ (red).
     (a) $j_x=1.36\times10^{10}\text{Am}^{-2}$, showing the transient motion
     in the pinned phase (PP).
     (b) $j_x=2.53\times10^{10}\text{Am}^{-2}$, where the skyrmion is transported
     with constant average velocity along $-x$.
     (c) $j_x=3.31\times10^{10}\text{Am}^{-2}$,
     corresponding to the transition 
     between the constant velocity phase (CVP) and the reentrant pinning phase (RPP).
     (d) $j_x=3.89\times10^{10}\text{Am}^{-2}$,
     showing the RPP with a large stabilized orbit.
    }
    \label{fig3}
\end{figure}

In Fig.~\ref{fig2} we plot the time average skyrmion velocity $\left\langle v_x\right\rangle$ versus the ac amplitude $j_x$.
When $j_x<1.5 \times10^{10} \text{Am}^{-2}$, 
the average skyrmion velocity is nearly zero.
This is the initial pinned phase (PP), and the slight deviation of the
velocity from zero is produced by
the brief initial period of time during which the skyrmion
orbit adjusts to changes in the ac driving amplitude.
A representative skyrmion
trajectory showing this transient motion is illustrated in Fig.~\ref{fig3}(a).
As $j_x$ increases, the skyrmion orbit increases
in size and becomes more unstable. 
At $j_x = 1.36\times10^{10}\text{Am}^{-2}$, the skyrmion velocity increases
abruptly to $\left\langle v_x\right\rangle\approx-3$ ms$^{-1}$,
indicating that a depinning threshold has been crossed.
The net motion of the skyrmion along the $-x$ direction
in this phase is associated with an orbit
that is too large to be stabilized between the linear protrusions,
and therefore evolves into a translating orbit that
carries the skyrmion across the sample.
For $1.5\times 10^{10}\text{Am}^{-2} < j_x < 2.9\times10^{10}\text{Am}^{-2}$,
the skyrmion is in a constant velocity
phase (CVP) with a translating orbit
that gives $\left\langle v_x\right\rangle\approx-3$ ms$^{-1}$.
The skyrmion flow in the $-x$ direction occurs along the hard
substrate asymmetry direction,
and the skyrmion must overcome the divots of the array
in order to flow.
The skyrmion trajectory in the CVP is shown in Fig.~\ref{fig3}(b).
The constant velocity is maintained even as $j_x$ varies because the
skyrmion translates by exactly one plaquette during each ac drive
cycle, and the driving frequency is being held constant.
For $j_x > 3.0\times10^{10}\text{Am}^{-2}$,
the system transitions to a reentrant pinning phase (RPP).
The average skyrmion velocity in the RPP is again not exactly zero
due to the transient adjustment of the orbit each time $j_x$ is
modified.
As $j_x$ increases,
when the RPP arises a new stable localized orbit develops that is much
more elongated than the orbit in the PP, as illustrated in Fig.~\ref{fig3}(c)
and (d).
The portion of the orbit extending along the $-x$ direction becomes so
extended that the skyrmion gets trapped underneath one of the linear
defects and is no longer able to jump into a neighboring plaquette
during the $-x$ portion of the ac drive cycle.
This reentrant transition into a pinned state is not as sharp as the
depinning transition found at lower $j_x$ since the trapping process
at high $j_x$ is
more gradual than the escape process at low $j_x$.


For ac driving along the $x$ direction, we observe three distinctive phases
with well defined behavior. The CVP,
where the average skyrmion velocity is constant,
can be useful for spintronic devices
where precise control of the skyrmion motion is crucial.
In addition, the skyrmion motion
can be switched on or off by a fine adjustment in the
external ac drive magnitude across the depinning transition.
Although we performed simulations only for
$j_x \leq 4\times10^{10}\text{Am}^{-2}$, we expect that similar behavior
will occur for larger values of $j_x$.
In particular, as $j_x$ increases, the
size of the skyrmion orbit can continue to increase
and may become unstable again, leading to the reemergence of a translating
orbit.

\subsection{Influence of the frequency $\omega$}
The motion described previously showed three distinctive phases
at fixed $\omega$, with
well defined ranges of $j_x$ for each phase.
We next investigate how the frequency $\omega$ of the ac drive
affects the dynamics by varying it over the range
$0.51\times10^{7}\text{Hz}\leq\omega\leq10.13\times10^{7}\text{Hz}$
and using the same range of $j_x$ values from before.

\begin{figure}
\includegraphics[width=1.0\columnwidth]{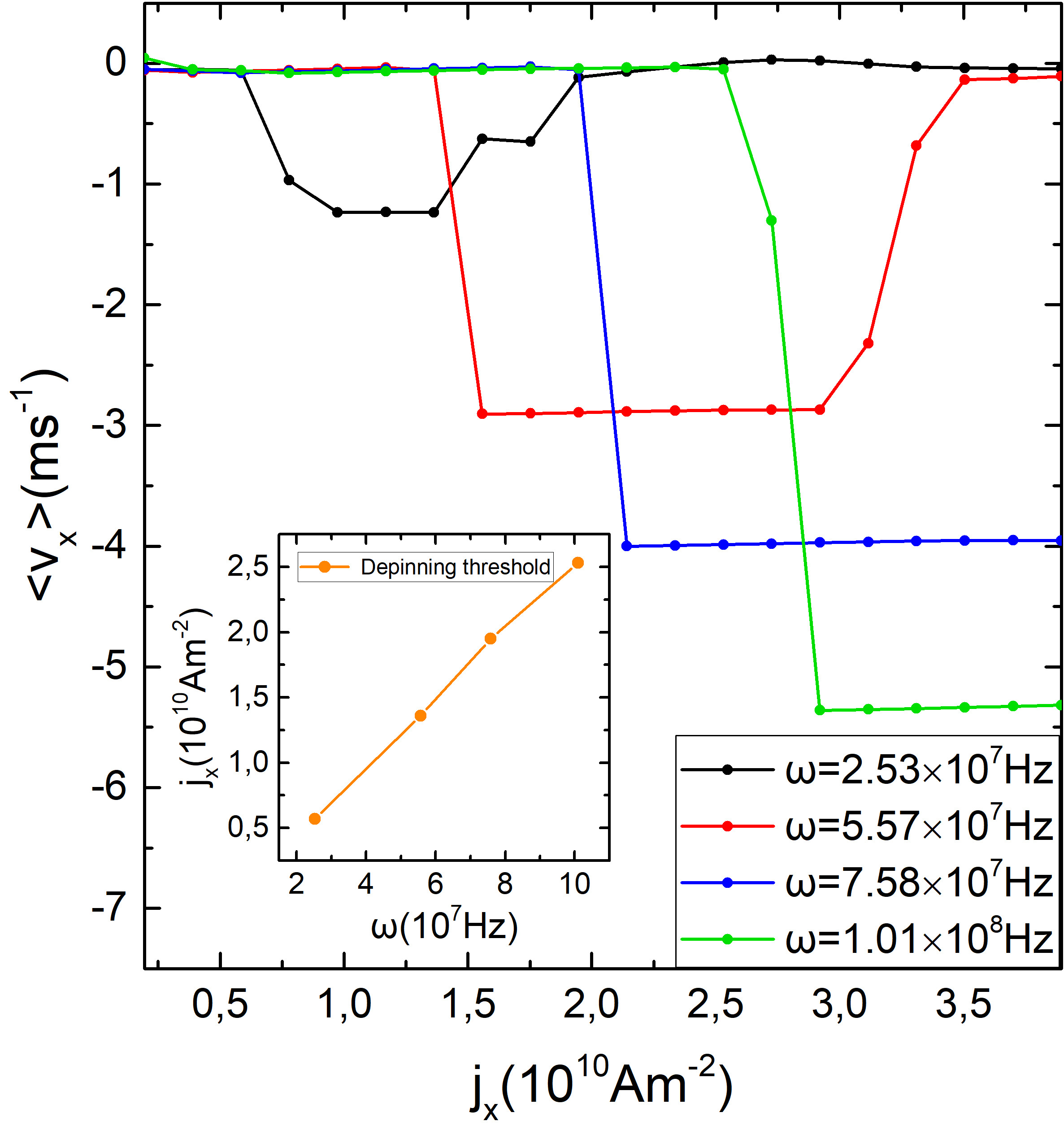}
\caption{$\left\langle v_x\right\rangle$ vs $j_x$
for the system from Fig.~\ref{fig1}
with ac driving applied along the $x$ direction
at different ac drive frequency
values of $\omega=2.53\times10^7$ Hz (black),
$\omega=5.57\times10^7$ Hz (red),
$\omega=7.58\times10^7$ Hz (blue), and
$\omega=1.01\times10^8$ Hz (green).
The pinned phase (PP), constant velocity phase (CVP),
and reentrant pinned phase (RPP)
appear for $\omega\leq 5.57\times10^7$Hz. For $\omega\geq 7.58\times10^7$Hz
the reentrant pinned phase does not occur over this range of $j_x$.
Inset: The value of $j_x$ at the depinning threshold marking the onset of
the CVP
vs $\omega$.}
\label{fig4}
\end{figure}

In Fig.~\ref{fig4} we plot the average skyrmion velocity
$\left\langle v_x\right\rangle$ versus $j_x$ 
for different values of $\omega$.
The behavior is similar in each case.
The PP appears
for all values of $\omega$. The CVP
is also observed for all values of $\omega$; however,
the width of this phase is strongly affected by $\omega$.
As $\omega$ increases, the onset of the CVP shifts towards higher values
of $j_x$, so that the range of the PP becomes larger.
This is more easily visible in the inset of Fig.~\ref{fig4}, where
we plot the depinning threshold versus $\omega$ and find a linear
dependence.
The increase in the depinning threshold occurs due to the oscillatory nature
of the ac drive. As $\omega$ increases, the oscillations in the driving
direction occur more rapidly, causing the skyrmion to experience a force in
any given direction for a shorter period of time, and shrinking the skyrmion
orbit accordingly.
Thus, a larger current amplitude must be applied at higher $\omega$ in order
to generate an orbit that is unstable enough to delocalize and produce
net dc motion.
Another interesting effect in
Fig.~\ref{fig4} is the increase in the
magnitude of the average skyrmion velocity in the CVP 
with increasing $\omega$.
Higher applied currents result in dc motion
with enhanced velocities.
Additionally, as $\omega$ increases,
the CVP extends up to higher values of $j_x$.
This is analogous to the increase of the depinning threshold as
a function of $\omega$;
for higher $\omega$ values, larger values of $j_x$ must be applied
in order to obtain a skyrmion orbit that is large enough to reach
the RPP.
For the highest value of $\omega$, $\omega= 7.58\times10^7$ Hz, the RPP
is not observed over the range of $j_x$ simulated here; however,
we expect that the RPP would appear at even higher 
values of $j_x$.

\begin{figure}
\includegraphics[width=\columnwidth]{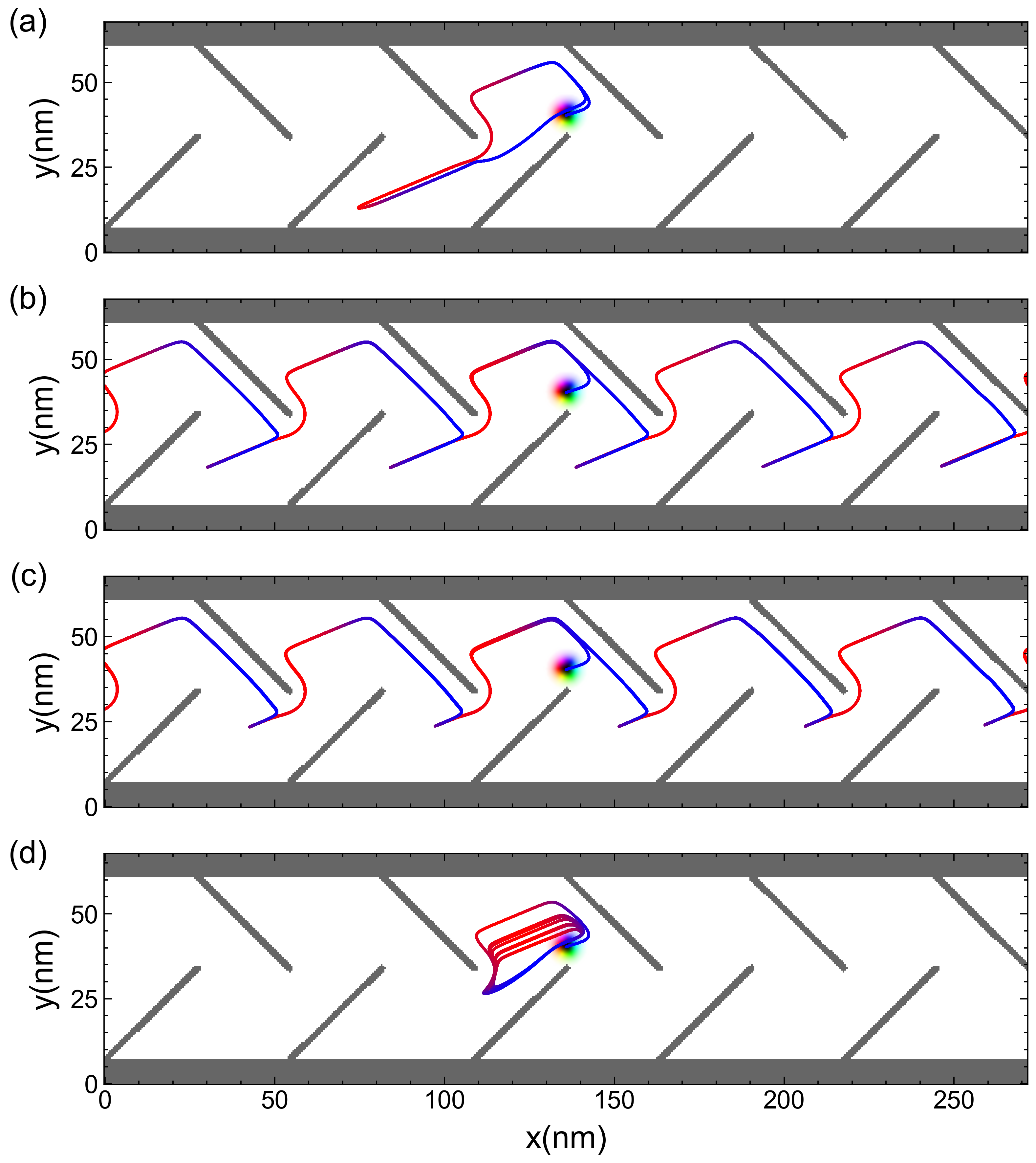}
\caption{Illustration of skyrmion trajectories for different values of $\omega$
with fixed $j_x=2.53\times10^{10}$Am$^{-2}$ for the system in
Fig.~\ref{fig4} with $x$ direction ac driving.
The skyrmion trajectory color indicates the phase of the ac
drive cycle as a gradient from 
positive $x$ (blue) to negative $x$ (red).
(a) $\omega=2.53\times10^7$ Hz, showing the reentrant pinned phase (RPP).
(b) $\omega=5.57\times10^7$ Hz, showing the constant velocity phase (CVP).
(c) $\omega=7.60\times10^7$ Hz in the CVP.
(d) $\omega=1.01\times10^8$ Hz, showing the pinned phase (PP).
}
    \label{fig5}
\end{figure}

Figure~\ref{fig5} illustrates the skyrmion trajectory
at different values of $\omega$.
In Fig.~\ref{fig5}(a) at $\omega=2.53\times10^7$ Hz, 
the skyrmion is trapped in the RPP.
As $\omega$ increases, the skyrmion orbit becomes narrower and
extends less far along the $x$ direction until the skyrmion
no longer experiences confinement underneath a linear defect. When this
occurs, the localized orbit
destabilizes, permitting the CVP to appear with
a translating orbit as shown in
Fig.~\ref{fig5}(b) for $\omega=5.57\times10^7$ Hz.
In Fig.~\ref{fig5}(c) at
$\omega=7.60\times10^7$ Hz,
the skyrmion is still in the CVP, but due to the increased
ac frequency, the dead end portion of the orbit extending in
the $-x$ direction becomes less pronounced.
When $\omega=1.01\times10^8$ Hz, the frequency is so high that the
orbit is no longer wide enough for the skyrmion to slip past the
linear defect into the next plaquette, and the PP emerges,
as shown in Fig.~\ref{fig5}(d).

\subsection{Conditions for skyrmion transport with ac drive along the $x$ direction}

\begin{figure}
\includegraphics[width=\columnwidth]{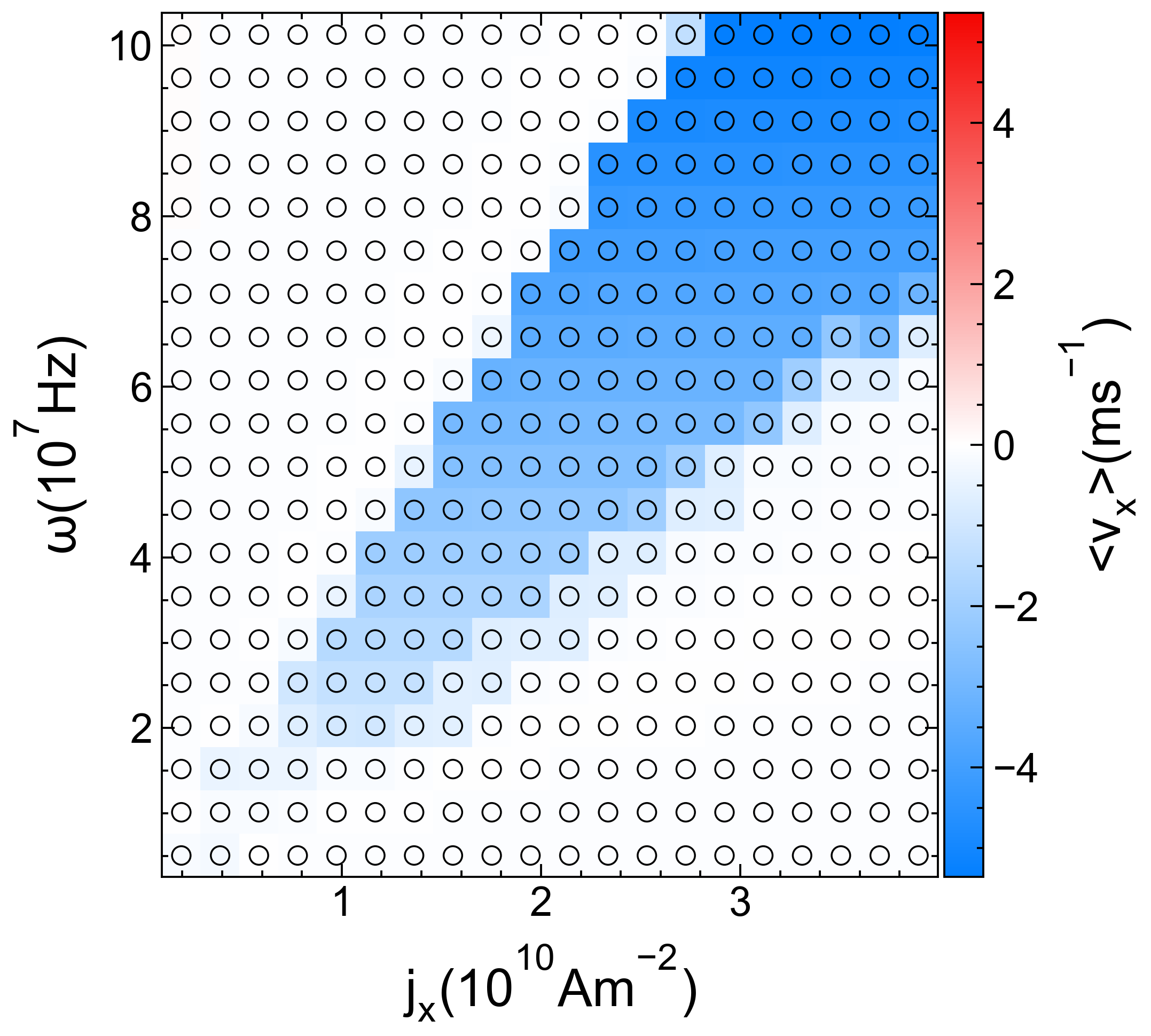}
\caption{Heatmap plot of $\left\langle v_x\right\rangle$ (color gradient)
 as a function of $\omega$ vs $j_x$ for the system in Fig.~\ref{fig1}
 with $x$ direction ac driving.
 In the blue regions skyrmion transport occurs
 along the negative $x$ direction, while in white regions
 the skyrmion is pinned. The circles indicate the discrete parameters
 used for the simulations.}
    \label{fig6}
\end{figure}

Now that we understand the individual effects of both ac drive magnitude, $j_x$, and the ac drive frequency, $\omega$, for ac driving in the $x$ direction,
we perform a series of simulations
in which we vary both parameters and identify
the optimal conditions for skyrmion
transport.
In Fig.~\ref{fig6} we plot a heatmap of the skyrmion average velocity
$\left\langle v_x\right\rangle$ as a function of $\omega$ versus
$j_x$.
Both the PP and RPP appear
as white regions
since both of these phases have very low velocities.
They are separated by the blue region, which represents the
CVP.
For values of $\omega$ and
$j_x$ falling above the blue region, the system is in the PP,
where the skyrmion exhibits no net motion and the trajectories are very
similar to those shown in Fig.~\ref{fig5}(d) and Fig.~\ref{fig3}(a).
For $\omega$ and $j_x$ values that are below the blue region,
the system is in the RPP, where the skyrmion
has no net motion and the trajectory is similar to
those shown in Fig.~\ref{fig5}(a)
and Fig.~\ref{fig3}(d).
In the CVP phase, the skyrmion trajectories are similar to those shown in
Fig.~\ref{fig5}(b,c) and Fig.~\ref{fig3}(b).
The speed at which the skyrmion is transported along the sample
in the CVP is governed
by $\omega$.
Note that as $\omega$ increases, the blue region becomes darker, indicating higher
velocity magnitudes.
Additionally, the onset of skyrmion transport and the range of $j_x$
values for which it occurs
also change as $\omega$ varies.

Figure~\ref{fig6} can be very useful for the design of a device using
a linear protrusion array of defects,
since it clearly shows the necessary parameters 
for skyrmion transport through the sample.
In addition, it indicates how rapidly
the skyrmion can be transported in the CVP.
In our simulations, the skyrmion is always stable and we did not observe
any annihilation effects, which is crucial for technological applications where skyrmions are to be used as
information carriers. We expect, however,
that if a much stronger ac drive amplitude is applied,
the skyrmion may be
annihilated when it comes into contact
with the linear protrusion magnetic walls.

\section{Ac drive along the $y$ direction}
In Section III we showed that
applying the ac drive along the $x$ direction may
induce skyrmion transport in the $-x$ direction.
Here, we investigate the same system from Fig.~\ref{fig1}
but apply the ac drive along the $y$ direction,
giving $j_x=0$ and $j_y \neq 0$.

\begin{figure}
    \includegraphics[width=0.8\columnwidth]{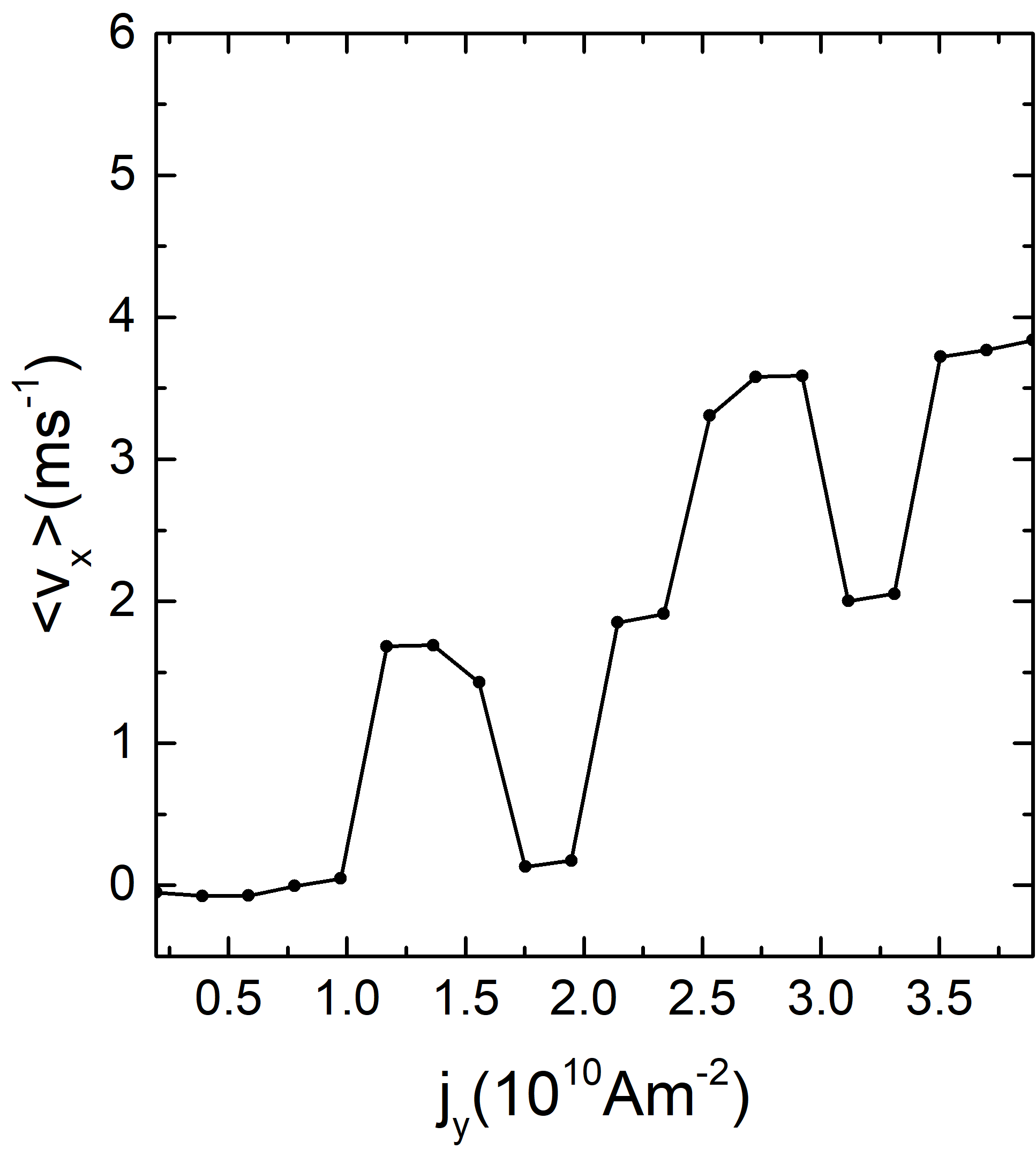}
    \caption{$\left\langle v_x\right\rangle$ vs
    $j_y$ with fixed $\omega=3.55\times10^7$ Hz for the system
      shown in Fig. \ref{fig1} with ac driving along the $y$ direction.
    Here the skyrmion moves in the positive $x$ direction.}
    \label{fig7}
\end{figure}

\begin{figure}
  \includegraphics[width=\columnwidth]{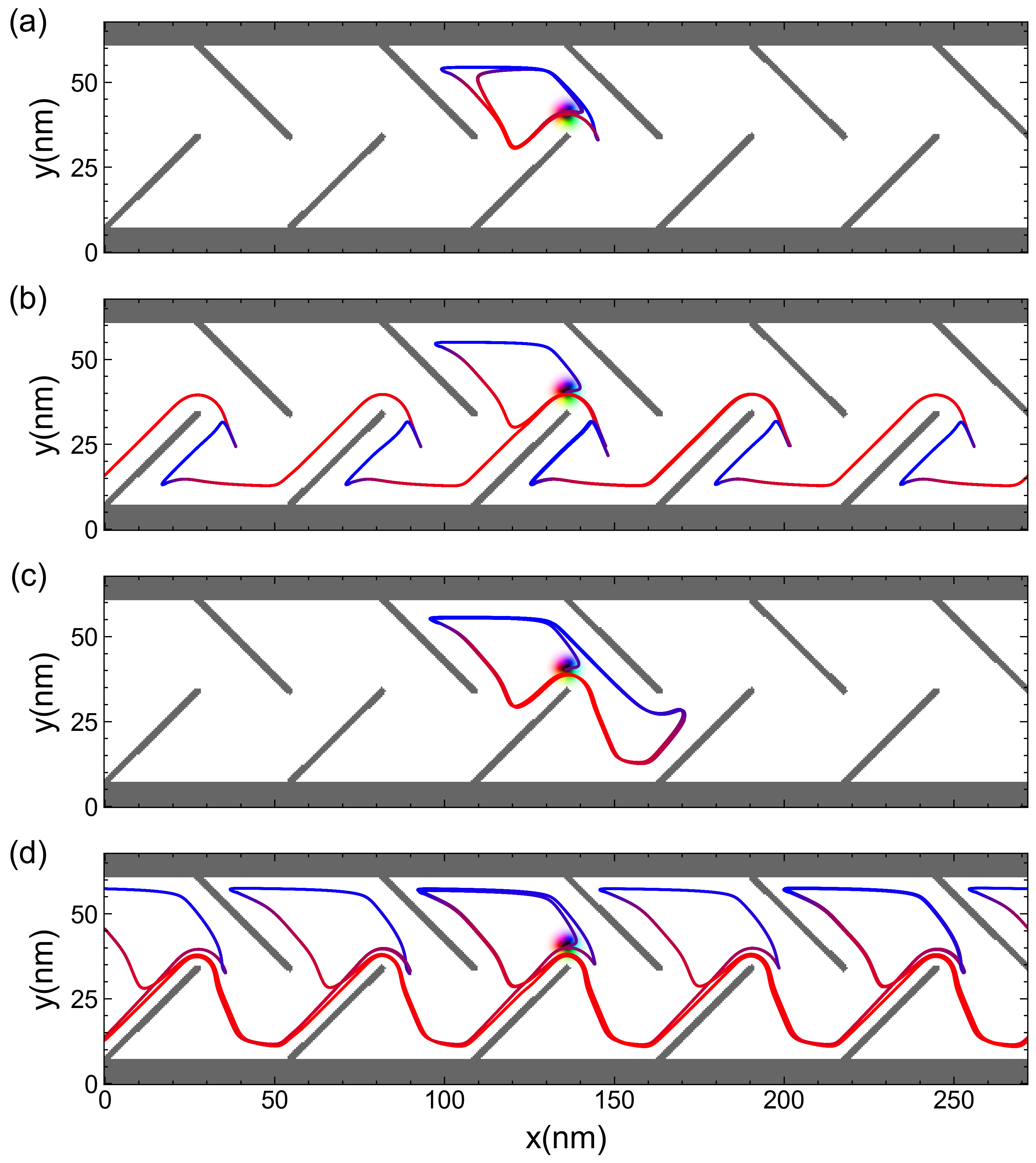}
\caption{Illustration of skyrmion trajectories for the system
in Fig.~\ref{fig1} under $y$ direction ac driving with fixed
$\omega=3.55\times10^7$ Hz.
The skyrmion trajectory color indicates the phase of the ac drive cycle as
a gradient from positive $y$ (blue) to negative $y$ (red).
(a) $j_y=0.97\times10^{10}$Am$^{-2}$ in the pinned phase (PP),
where the skyrmion is in a closed localized orbit.
(b) $j_y=1.36\times10^{10}$Am$^{-2}$ in the constant velocity phase (CVP),
where the skyrmion translates at
$\left\langle v_x\right\rangle\approx 1.68$ ms$^{-1}$
and moves one plaquette during every ac drive cycle.
(c) $j_y=1.95\times10^{10}$Am$^{-2}$ in the reentrant pinned phase (RPP),
where the skyrmion enters a stable localized orbit and the net dc motion
drops to zero.
(d) $j_y=3.69\times10^{10}$Am$^{-2}$ in a second CVP,
where the skyrmion translates at
$\left\langle v_x\right\rangle\approx 3.8$ ms$^{-1}$
and moves two plaquettes during every ac drive cycle.}
    \label{fig8}
\end{figure}

In Fig. \ref{fig7} we plot the skyrmion average velocity $\left\langle v_x\right\rangle$
as a function of the ac drive amplitude $j_y$ for fixed
$\omega=3.55\times10^7$ Hz.
We find
several different dynamic regimes
that we can classify as a pinned phase (PP),
a reentrant pinned phase (RPP), and
different types of constant velocity phases (CVPs).
For $j_y\leq0.97\times10^{10}$Am$^{-2}$, the skyrmion exhibits no
net motion and is in the PP as illustrated in Fig.~\ref{fig8}(a).
As was the case for $j_x \neq 0$ driving,
we find that the velocities are not
exactly zero since the skyrmion
undergoes transient motion before the localized orbit
becomes stabilized.
In the interval
$1.75\times10^{10}\text{Am}^{-2}\leq j_y\leq2.34\times10^{10}\text{Am}^{-2}$,
the skyrmion orbit is unstable and the dc velocity takes the value
$\left\langle v_x\right\rangle\approx 1.68$ ms$^{-1}$,
indicating that the skyrmion is in the CVP
illustrated in Fig.~\ref{fig8}(b).
Here the skyrmion translates by one plaquette during every ac drive
cycle.
The skyrmion has a greater interaction with the lower part of the sample
when the ac driving is in the $y$ direction due to the
skyrmion Hall angle effect,
and the net transport is in the positive $x$ direction, unlike the case
of $x$ direction ac driving which produced dc motion in the negative $x$
direction.
Above
$j_y = 1.36\times10^{10}\text{Am}^{-2}$,
the skyrmion average velocity drops to a value that is very close to zero.
In this region, a localized skyrmion orbit becomes stable again and the
dc motion is lost, giving a
RPP as shown in Fig.~\ref{fig8}(c).
Here the width of the skyrmion orbit matches the spacing between
the linear protrusions.
When
$j_y = 1.94\times10^{10}\text{Am}^{-2}$, the localized orbit destabilizes
because it is now too wide to fit inside a single plaquette,
and dc motion reappears.
Notice that the skyrmion is now translating by two plaquettes during every dc
drive cycle. A blue orbit appears in the top portion of every plaquette only
because the total number of plaquettes is odd. Figure~\ref{fig7} shows that the
skyrmion velocity for the state in Fig.~\ref{fig8}(d) is twice as large
as that in Fig.~\ref{fig8}(b).
For larger values of $j_y$, the skyrmion dynamics oscillate among
a series of CVPs where the velocity locks to a constant value that differs
from one CVP to the next and is determined by the number of plaquettes
the skyrmion can translate in each ac drive cycle.
In every CVP, the skyrmion follows a distinct
delocalized orbit as it translates through the sample.
Figure~\ref{fig8}(d) 
shows an example of the skyrmion trajectory
in the CVP at $j_y=3.69\times10^{10}$Am$^{-2}$.
The fact that the skyrmion translates along the easy direction of the
substrate asymmetry
for $y$ direction driving
makes it possible for the skyrmion to move through multiple
plaquettes per ac drive cycle,
giving multiple CVP states. This is in contrast to the single CVP state found
for $x$ direction driving, when the motion is along the hard
direction of the substrate asymmetry and the translating orbit is 
confined by the substrate in such a way that the skyrmion can travel
only exactly one plaquette during each ac drive cycle.

\subsection{Influence of the frequency $\omega$}

\begin{figure}
    \includegraphics[width=0.8\columnwidth]{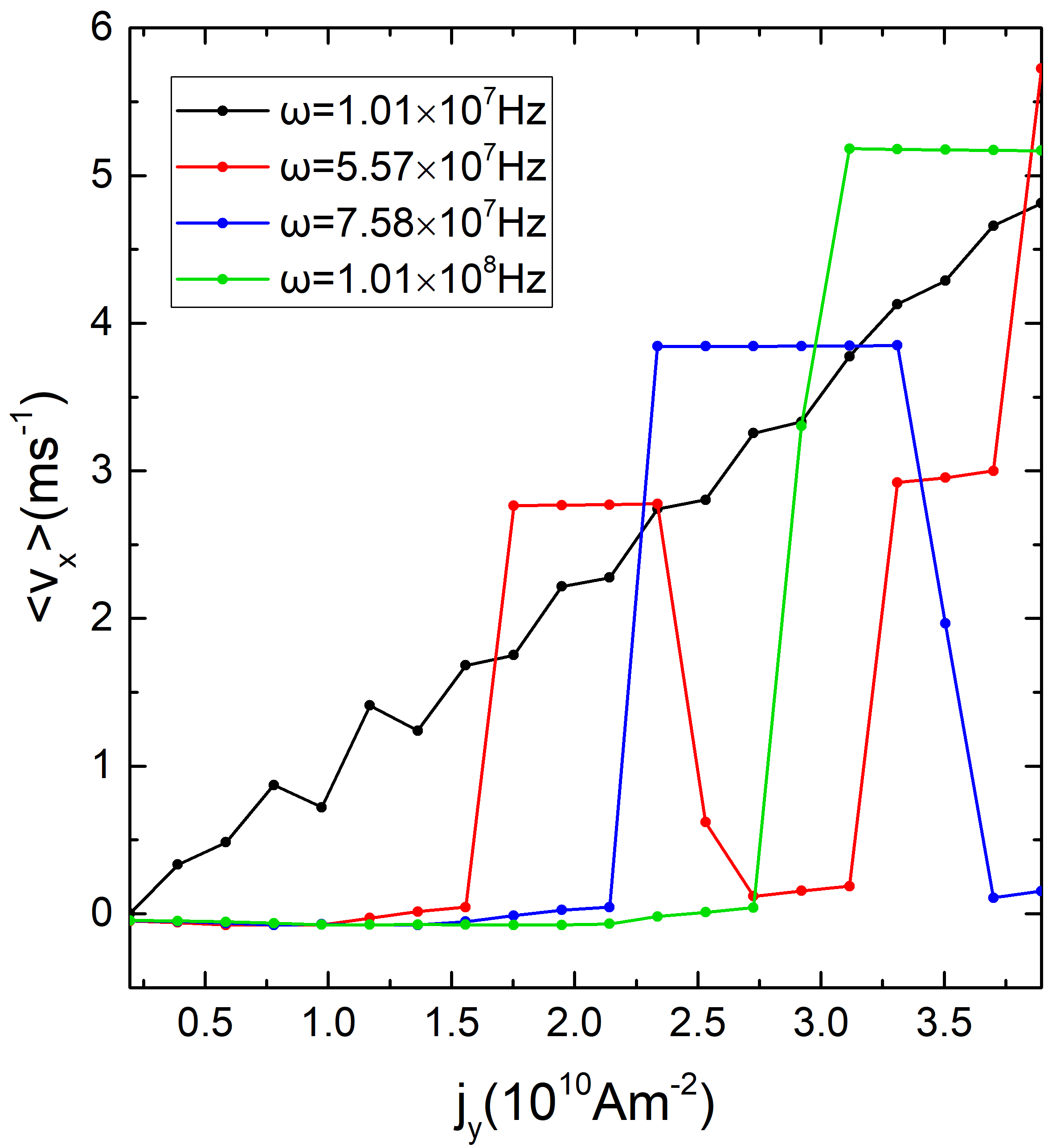}
    \caption{$\left\langle v_x\right\rangle$ vs
      $j_y$ for the system from Fig.~\ref{fig7} with ac driving
      along the $y$ direction at
    $\omega=1.01\times10^7$ Hz (black), $\omega=5.57\times10^7$ Hz (red),
    $\omega=7.58\times10^7$ Hz (blue), and $\omega=1.01\times10^8$ Hz (green).
}
    \label{fig9}
\end{figure}

\begin{figure}
  \includegraphics[width=\columnwidth]{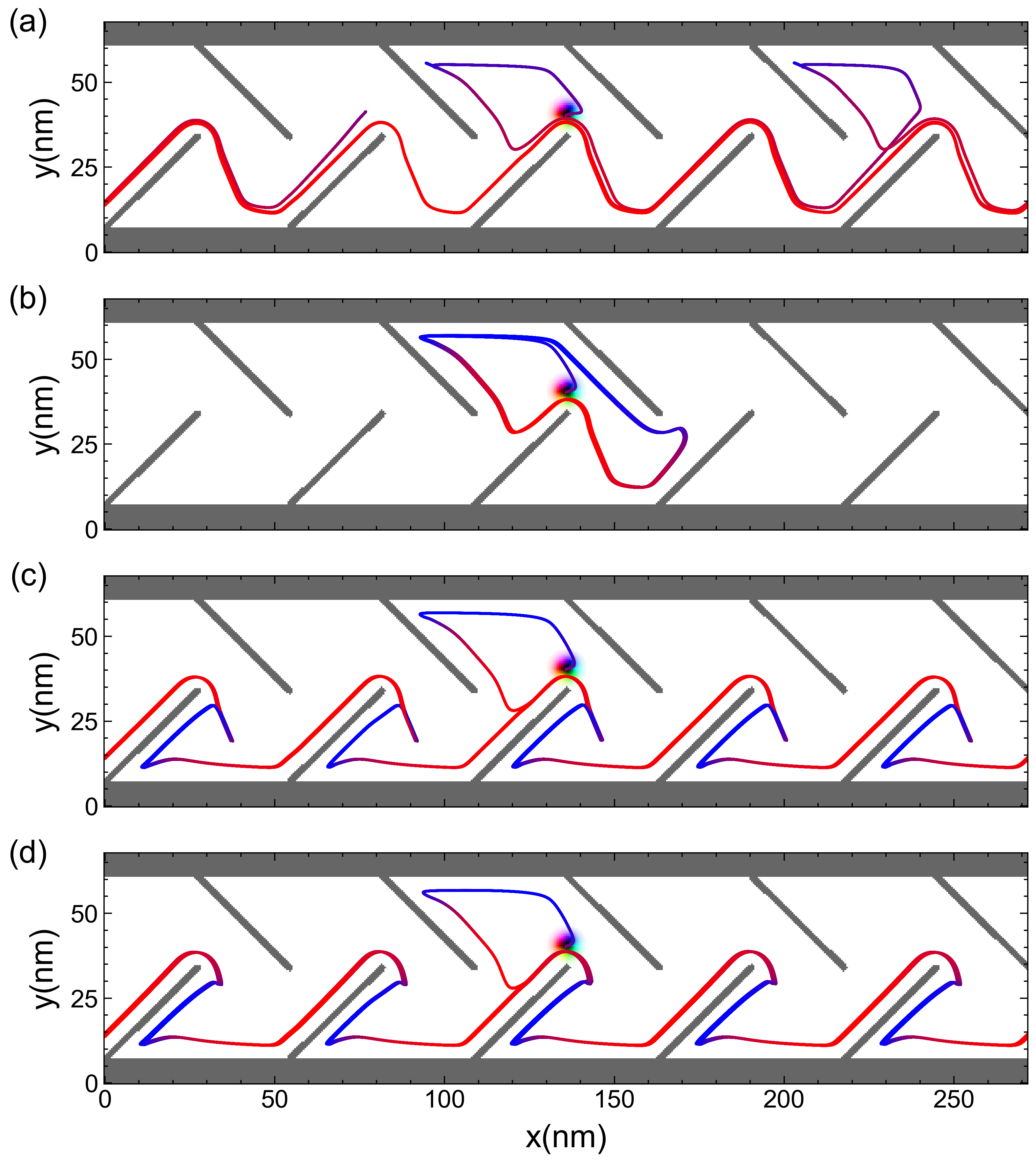}
  \caption{Illustration of skyrmion trajectories for the system in
   Fig.~\ref{fig9} under $y$ direction ac driving
   with fixed $j_y=3.12\times10^{10}$ Am$^{-2}$. The skyrmion trajectory color
   indicates the phase of the ac drive cycle as a gradient from positive
   $y$ (blue) to negative $y$ (red).
  (a) $\omega=1.01\times10^7$ Hz.
  (b) $\omega=5.57\times10^7$ Hz.
  (c) $\omega=7.58\times10^7$ Hz.
  (d) $\omega=1.01\times10^8$ Hz.}
    \label{fig10}
\end{figure}

We next consider
how the ac driving frequency $\omega$ affects the dynamics
under $y$ direction driving
by varying $\omega$ over the 
range
$0.51\times10^{7}\text{Hz}\leq\omega\leq10.13\times10^{7}\text{Hz}$.
In Fig.~\ref{fig9} we plot the average skyrmion velocity $\left\langle v_x\right\rangle$ versus $j_y$ for selected values of $\omega$.
When
$\omega=1.01\times10^7$ Hz, the CVP and RPP vanish.
As low values of $\omega$ such as this one,
the skyrmion can coast along the bottom of the array over a distance
of many plaquettes during the $-y$ portion of the ac drive
cycle, as shown in Fig.~\ref{fig10}(a) at $j_y=3.12\times10^{10}$ Am$^{-2}$.
The Magnus term rotates the repulsive force from the protrusions and walls
into a $+x$ velocity, resulting in a Magnus velocity boost effect.
The confined excursion into the upper portion of the sample provides
only a small perturbation to the boosted $+x$ velocity and is insufficient
to quantize the velocity values.
As a result, $\left\langle v_x\right\rangle$
increases almost linearly with the ac drive amplitude $j_y$.
For $\omega=5.57\times10^7$ Hz,
$\omega=7.58\times10^7$ Hz, and $\omega=1.01\times10^8$ Hz,
the behavior is similar to what was observed in
in Fig.~\ref{fig7},
with the CVP and RPP both present but extending over a
wider range of $j_y$ values compared to the system with
$\omega=3.55\times10^7$ Hz.
For $\omega=5.57\times10^7$ Hz
the skyrmion becomes trapped in a RPP, as
illustrated in Fig.~\ref{fig10}(b) at
$j_y=3.12\times 10^{10}$ Am$^{-2}$.
Here the size of the skrymion orbit matches the length scale
of the plaquette.
For $\omega=7.58\times10^7$ Hz at the same
value of $j_y$, the orbit is delocalized again
and the skyrmion flows along the positive $x$ direction as shown
in Fig.~\ref{fig10}(c).
Figure~\ref{fig9} indicates that
the skyrmion average velocity in each CVP becomes larger
for higher frequencies, 
as illustrated
in Fig.~\ref{fig10}(d) where we show
another type of CVP with higher average
velocity at $\omega=1.01\times 10^8$ Hz.
The velocity increase
is simply due to the higher ac driving frequency; the skyrmion
still translates by one plaquette per ac drive cycle but the
drive cycle is shorter at higher $\omega$, so the skyrmion
moves faster.
The onset of the CVP
in Fig.~\ref{fig9} shifts to higher $j_y$ with increasing
$\omega$ because
at high ac driving frequencies, the 
skyrmion orbits become narrower and so larger values of
$j_y$ must be applied to destabilize the localized orbits.

\subsection{Conditions for skyrmion transport with ac drive along the $y$ direction}

\begin{figure}
  \includegraphics[width=\columnwidth]{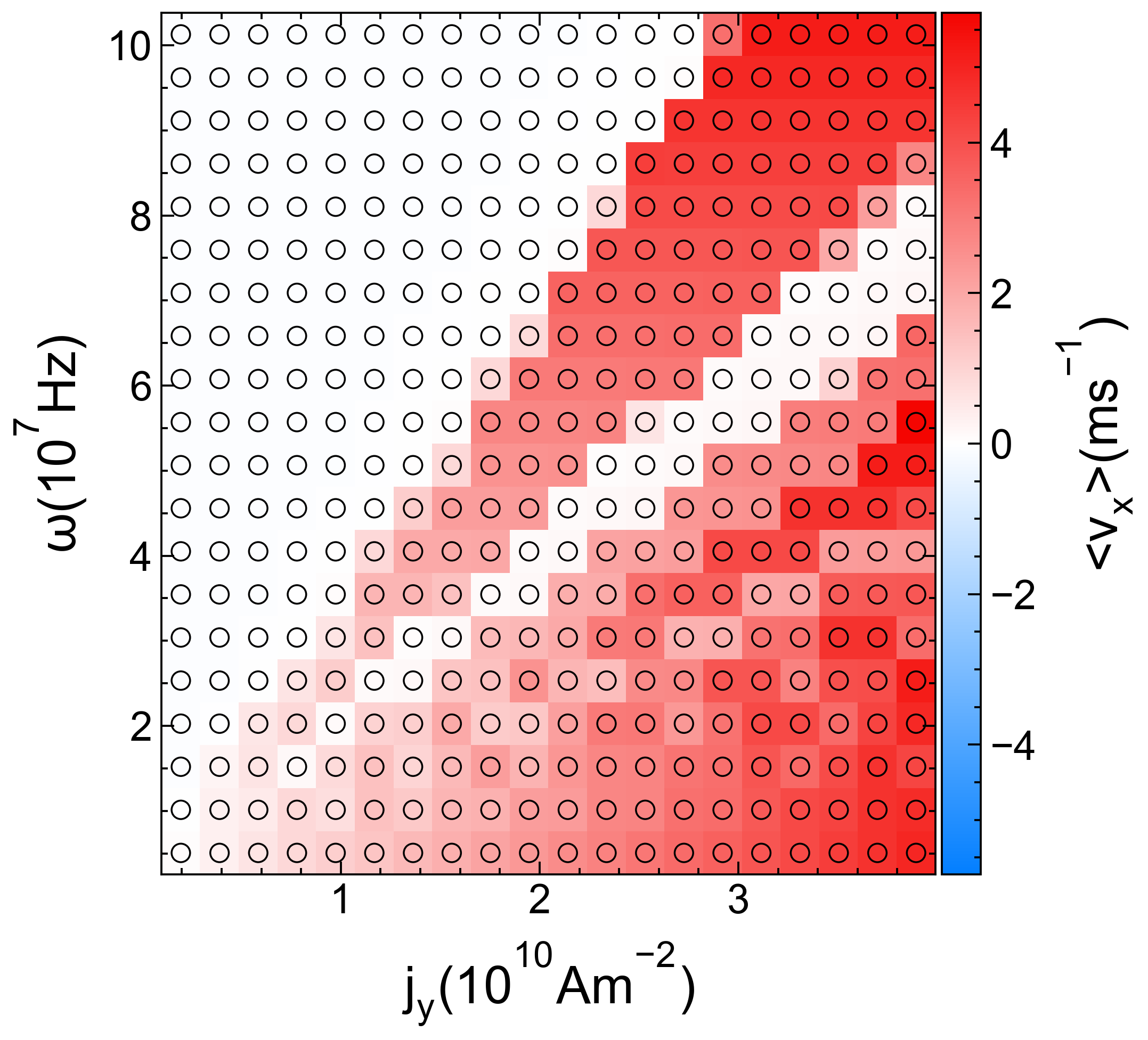}
  \caption{Heatmap plot of $\left\langle v_x\right\rangle$ (color gradient)
   as a function of $\omega$ vs $j_y$ for the system in
   Fig.~\ref{fig1} with $y$ direction ac driving.
   In the red regions skyrmion transport occurs along the positive $x$
   direction,
   while in white regions
   the skyrmion is pinned.
   The circles indicate the discrete parameters used for the
   simulations.}
    \label{fig11}
\end{figure}

In Fig.~\ref{fig11} we plot a heatmap of the skyrmion average velocity
$\left\langle v_x\right\rangle$ as a function of 
the frequency $\omega$
versus
the drive magnitude $j_y$.
There are two large red regions corresponding to the
dynamic phases where the skyrmion is being transported. 
The white regions correspond to phases where the skyrmion
exhibits no net motion.
The first white region that occurs for lower values of $j_y$ is
the PP
where the localized skyrmion orbit is stabilized with no net motion.
At slightly larger values of $j_y$ we find
the first red region, corresponding to the CVP
in which the localized skyrmion orbit is unstable and
the skyrmion translates through the sample along the
positive $x$ direction.
In this phase, as $\omega$ increases, the range of $j_y$
where the CVP is stable increases,
and the average skyrmion velocity also increases
in a manner analogous to that discussed
for Fig.~\ref{fig6}.
The second white region, located between two red regions,
is the RPP,
where the increase in $j_y$ stabilizes the previously unstable
localized skyrmion orbit, resulting in a state with
$\left\langle v_x\right\rangle = 0$.
In addition, as $\omega$ increases, the RPP expands over a wider
range of $j_y$.
The second red region that occurs for larger values of $j_y$
corresponds to a mixture of CVP and a
skyrmion transport phase in which the velocity
increases almost linearly with $j_y$.
In this region, for larger values of $j_y$
the system is in a clearly defined
CVP phase where multiple constant velocity states can be accessed
by varying the ac drive amplitude.
As $\omega$ is reduced, the CVP is lost
and the skyrmion velocity increases linearly with $j_y$.
This is a result of the velocity boost effect from the Magnus
term, which makes the motion during the $+y$ and $-y$ portions
of the ac drive cycle
very asymmetric and destroys the velocity quantization found for
larger values of $\omega$.

As we have seen, when the ac drive is applied along the $x$ direction,
the skyrmion flows along $-x$, the hard substrate asymmetry direction,
while when the ac drive is applied along the $y$ direction,
the skyrmion instead flows along $+x$, the easy substrate asymmetry
direction.
The plots in Figs.~\ref{fig6} and \ref{fig11}
show in great detail the possibilities for achieving
controlled skyrmion motion by tuning the values of
$\omega$ and either $j_x$ or $j_y$.
We note that for the parameters we consider,
no annihilation effects were observed; however, for sufficiently
large ac drive amplitudes, we expect that the skyrmion would be pushed
hard enough against the walls and linear protrusions that annihilation
events would occur. Thus there is a finite range of ac driving where
control of stable skyrmion motion can be achieved.

\section{Summary}
Using an atomistic model for simulating individual
atomic magnetic moments, we investigated the dynamical
behavior of a single skyrmion interacting
with a linear protrusion array of defects with
ac driving applied along
either the $x$ or $y$ directions in the absence of thermal effects.
When the ac driving is parallel to the $x$ direction, the
skyrmion can be transported along the negative $x$ 
or hard substrate asymmetry direction over a range
of ac drive amplitudes. There are three distinct phases of motion.
At low ac amplitudes, the skyrmion enters a stable localized orbit
and has no net dc motion, giving a pinned phase.
When the ac amplitude increases, the
localized skyrmion orbit expands in size
and becomes unstable, leading to the emergence of
a constant velocity phase
in which the skyrmion follows a
translating orbit along the negative $x$ direction.
As the ac driving amplitude becomes even larger, the skyrmion orbit
continues to grow until a localized
orbit restabilizes again due to confinement by the linear protrusions,
resulting in a reentrant pinned phase with no net dc transport.
When the ac drive frequency increases,
we find that the range of ac drive amplitudes over which each
phase is stable
increases,
and that the skyrmion average velocity in the constant velocity phase also
increases.

When the ac drive is applied along the $y$ direction, the skyrmion
can be transported along the positive $x$ or easy flow direction of
the substrate asymmetry
for a range of
ac amplitudes.
The velocity boost effect produced by the Magnus term for $y$ direction
driving in the presence of the substrate permits
a greater range of translating orbits to appear, and
we observe a richer dynamical response compared to driving along the $x$
direction.
At low values of ac drive amplitudes, the skyrmion orbit is localized
and we find a pinned phase with no net motion.
As the ac amplitude increases, the localized orbit destabilizes and
the skyrmion enters a constant velocity phase in which it translates along the
positive $x$ direction with constant velocity.
Further increases in the ac drive amplitude restabilize a
localized orbit, resulting in the emergence of a reentrant pinned phase
with no net motion.
For sufficiently high values of the ac drive amplitude,
the skyrmion orbit becomes too large to
localize, and we find a series
of distinct constant velocity phases,
each of which has a different average velocity value
determined by the number of plaquettes the skyrmion traverses
during each ac drive cycle.
Just as in the case for $x$ direction ac driving, increasing the ac driving
frequency for $y$ direction ac driving causes each phase to extend over a
wider range of ac drive amplitudes and also increases the average
velocity in the constant velocity phase.
Above the reentrant pinning phase, the constant velocity phases become
more stable as the ac frequency increases, while for small frequencies
these phases disappear completely and are replaced by a regime in which
the average skyrmion velocity increases linearly with
increasing ac drive amplitude.

Our findings can be useful
for realizing novel spintronic devices where controlled skyrmion motion is crucial.
In our system, the ability to control the direction and velocity
of the skyrmion motion makes it possible to efficiently transport
information in a device where the skyrmion serves as an information
carrier.
The level of skyrmion control that we obtain in our sample is similar to
the control achieved in Ref.~\cite{souza_skyrmion_2021}, where the skyrmion
is also controlled precisely using ac currents. 
The key difference from the previous work
is that in the linear protrusion device considered here,
the skyrmion average velocities are similar for transport in both the
positive and negative $x$ directions,
resulting in a more energy-efficiency skyrmion transport overall.

\section*{Acknowledgments}
This work was supported by the US Department of Energy through the Los Alamos National Laboratory. Los
Alamos National Laboratory is operated by Triad National Security, LLC, for the National Nuclear Security
Administration of the U. S. Department of Energy (Contract No. 892333218NCA000001). 
\\
J.C.B.S acknowledges funding from Fundação de Amparo à Pesquisa do Estado de São Paulo - FAPESP (Grant 2022/14053-8).
\\
We would like to thank Dr. Felipe F. Fanchini for providing the computational resources used in this work. 
These resources were funded by the Fundação de Amparo à Pesquisa do Estado de São Paulo - FAPESP (Grant: 2021/04655-8).

\bibliography{mybib}

\end{document}